\documentclass[aps,prl,reprint,twocolumn,amsmath,amssymb]{revtex4-1} 
\pdfoutput=1

\usepackage{amsmath,amsthm}
\usepackage{graphicx}
\usepackage{dcolumn}
\usepackage{bm}
\usepackage{color}
\usepackage[english]{babel}
\usepackage{epsfig}
\usepackage{multirow}
\usepackage{siunitx}
\usepackage[disable]{todonotes}

\graphicspath{{./figures/}{./sketches/}}

\newcommand{\Vect}[1]{\ensuremath{\bm{#1}}} 

\newcommand{\ie}{i.~e.~}
\newcommand{\etal}{et~al.~}

\begin{document}

\preprint{Preprint}

\title{Importance of Varying Permittivity on the Conductivity of Polyelectrolyte Solutions}

\author{Florian \surname{Fahrenberger}}
\author{Owen A. \surname{Hickey}}
\author{Jens \surname{Smiatek}}
\author{Christian \surname{Holm}} \email{holm@icp.uni-stuttgart.de}

\affiliation{Institut f\"{u}r Computerphysik, Universit\"{a}t Stuttgart, Allmandring 3, Stuttgart 70569, Germany}


\date{\today}

\begin{abstract}
 Dissolved ions can alter the local permittivity of water, nevertheless most theories and simulations ignore this fact. We present a novel algorithm for treating spatial and temporal variations in the permittivity and use it to measure the equivalent conductivity of a salt-free polyelectrolyte solution. Our new approach quantitatively reproduces experimental results unlike simulations with a constant permittivity that even qualitatively fail to describe the data. We can relate this success to a change in the ion distribution close to the polymer due to the built-up of a permittivity gradient.

\end{abstract}
\maketitle

The dielectric permittivity $\varepsilon$ measures the polarizability of a medium subjected to an electric field and is one of only two fundamental constants in Maxwell's equations. The relative permittivity of pure water at room temperature is roughly 78.5, but charged objects dissolved in the fluid significantly reduce the local dielectric constant because water dipoles align with the local electric field created by the object rather than the external field~\cite{hasted48a,bonthuis11a,sega13a}. When ions accumulate in the vicinity of a charged object they further reduce the local dielectric constant~\cite{hess06a,bonthuis11a,bonthuis12a}, causing a permittivity gradient that repels ions from the surface~\cite{fahrenberger13c,ma14a,lamm97a,nakamura13a}. The screening of electrostatic forces between charged objects is therefore affected, and it also influences any properties that depend on the specific ion distribution. However, almost all computational and theoretical work to date using an implicit water model assumes a constant dielectric permittivity. This is partially due to a lack of suitable numerical approaches, since only few electrostatic algorithms can include spatial changes in the dielectric permittivity~\cite{tyagi10a,jadhao12a,jadhao13a,wang15a}.

Electrophoresis is the directed motion of an object in an aqueous solution subject to an external electric field. The relative ease with which electric fields can be applied experimentally has led to wide use of electrophoresis in the characterization of polymers~\cite{viovy00a,dorfman10a,shendruk12a}, colloids~\cite{russel89a,wiersema66a,obrien78a,lobaskin07a,lobaskin08a}, and cells~\cite{hayashi01a,duval10a}, which tend to ionize in aqueous solutions. Measuring the electrophoretic velocity of individual particles is often difficult from a technical standpoint. For this reason, polyelectrolyte solutions are often characterized by their conductivity~\cite{cametti14a,dobryin95a,liparostir09a,colby97a,kwak75a,manning69a,schmitt73a,bordi05a,bordi04a,fischer08a}.

In both electrophoresis and conductivity, the distribution of oppositely charged counterions around the object determines the magnitude of the relative velocity between the object and the fluid. The surrounding counterion cloud is comprised of two separate layers: the Stern layer and the Debye layer~\cite{ohshima06b}. The Stern layer, often called the stagnant layer, consists of strongly adsorbed ions adjacent to the charged object that reduce the effective surface charge. Beyond the Stern layer is the Debye layer, also called the diffuse layer, which consists of ions that are free to move relative to the charged surface. The two layers are collectively referred to as the electric double layer (EDL).

For the study of charged macromolecules a coarse-grained approach is necessary due to the excessive system size.  Molecular Dynamics (MD) simulations of charged systems typically work with the restricted primitive model by simulating the ions as hard spheres while accounting for the solvent implicitly through a constant background dielectric. Crucially, the solvent mediates hydrodynamic interactions and reduces the electrostatic interactions due to its polarizability. Hydrodynamic interactions significantly impact the electrokinetic properties and lead to qualitatively different behavior~\cite{grass08a,grass09a}. Polarizability actually depends on the local electric field~\cite{gongadze12a,walker11a}: these local variations in the polarizability thicken the Debye layer by repelling the counterions~\cite{fahrenberger13c,ma14a,zwanikken13a,lamm97a,curtis15a,wang15a}. More detailed atomistic computer simulations and experiments have also indicated pronounced deviations to standard theories for ion distributions around charged objects if the continuum solvent approach  
is replaced by an explicit water environment~\cite{hess06a,savelyev06a,heyda12a,paterova13a,netz12a,kunz10a,lo12a,smiatek14a,smiatek14b,wohlfarth15a}. Hence, these findings show that the molecular properties of water are important for a detailed description of the EDL. However, the consequences for electrokinetic properties are largely unknown.

In this letter, we show that including spatially and temporally varying dielectric properties significantly influences the structure of the Debye layer and the dynamic properties of polyelectrolytes. By means of a novel algorithm, we locally couple the permittivity to the local ion concentration. When applied to a polyelectrolyte solution in an external electric field, the algorithm quantitatively reproduces experimental data for the conductivity, while simulations assuming a constant dielectric background disagree qualitatively with experiment. We
show that the decreased permittivity in the vicinity
of the polyelectrolyte reduces the fraction of condensed
counterions on the polyelectrolyte backbone, explaining the
 experimentally observed increase in the equivalent conductivity at
high monomer concentrations.

\begin{figure}[!htbp]
	\centering
	\includegraphics[width=0.8\linewidth]{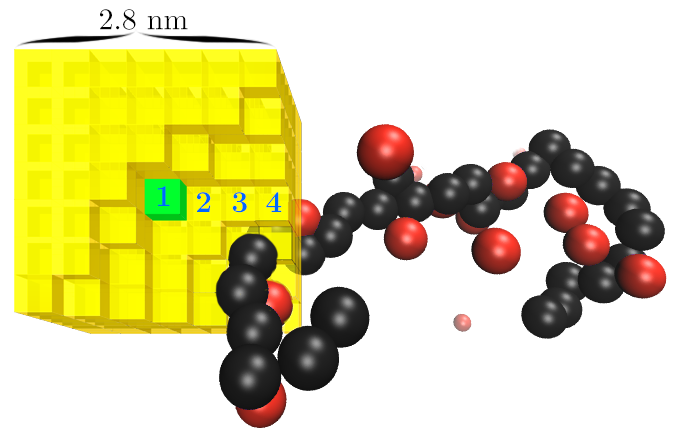}
	\caption{(color online) Our system setup with polyelectrolytes (black spheres) and counterions (red spheres). The scheme to calculate local charge concentrations is depicted by the lattice in the top left corner. The concentration of a lattice cell is determined via a weighted summation of all charged particles in the $7^3$ surrounding MEMD lattice cells. Ions thus influence the local permittivity within a distance $d=\SI{1.4}{nm}$, or two Bjerrum lengths.}
	\label{fig:system-setup}
\end{figure}
We performed standard coarse-grained simulations using a Weeks-Chandler-Anderson (WCA) potential~\cite{weeks71a} for steric interactions with an equilibrium distance of $\sigma=\SI{0.3}{nm}$ between particles. Adjacent monomers are connected by nitely extensible nonlinear elastic
(FENE) bonds. Simulations looking
at static ion distributions were performed with a Langevin thermostat, while simulations in which dynamic quantities were measured used the D3Q19 lattice-Boltzmann (LB) fluid in ESPResSo~\cite{rohm12a,arnold13a,limbach06a} and the corresponding thermostat. A detailed description of the simulation method can be found in the supplemental material\footnote{See Supplemental Material url}.

The interactions between charges are calculated using an extension of a local electrostatics algorithm that was first introduced by A.~Maggs~\cite{maggs02a} and later independently adopted for MD simulations by J.~Rottler~\cite{rottler04a} and I.~Pasichnyk~\cite{pasichnyk04a}. This method of calculating electrostatic interactions has been coined
Maxwell Equations Molecular Dynamics (MEMD). 
This algorithm's locality permits arbitrary changes in the dielectric constant~\cite{fahrenberger14a}. The motion of charges $q_i\Vect{v}_i$ is interpolated to a regular lattice. The resulting electric current gives rise to a change in the magnetic field $\Vect{B}$ and the displacement field $\Vect{D} = \varepsilon \Vect{E}$, following Maxwell's equations. The algorithm, however, treats the propagation speed of the magnetic field $c$ as a tunable parameter. For a wide range of values of $c$, MEMD produces the correct particle dynamics and statistic observables~\cite{maggs02a,fahrenberger14a}. The algorithm is therefore computationally efficient, since only two update steps for the electromagnetic fields are performed per MD time step. The permittivity $\varepsilon$ is used twice: The coupling of the displacement field to the magnetic field and vice versa (Amp\`{e}re's law and Faraday's law). Effectively, this creates two new driving forces if $\varepsilon$ depends on space and time: The influence of magnetic waves that are partially reflected in lattice cells with variable permittivity, and a force directly pointing in direction of the permittivity gradient, following the Lorentz force $\Vect{F}_L = q\Vect{D}/\varepsilon + \Vect{v}\times\Vect{B}$.

The MEMD grid spacing is $a_\mathrm{MEMD}=\SI{0.4}{nm}$ and the time step is set to $\Delta t_\mathrm{MEMD} = \Delta t_\mathrm{MD}$. The last parameter is an artificial mass $f_\mathrm{mass} = 0.05 m_0$. To take into account changes in the solvent polarization due to ions in the double layer~\cite{hess06a,bonthuis11a,bonthuis12a}, the local dielectric permittivity is calculated from the nearby ion density using the empirical function found by B.~Hess~\etal{}~\cite{hess06a},
\begin{equation}
	\varepsilon = \frac{78.5}{1+0.278\cdot C} ,
	\label{eq:salt-map}
\end{equation}
where $C$ is the molar salt concentration in [mol/L] or [M].
For each lattice cell, the charge density is averaged in the surrounding $7^3$ cells as shown in the top left of Figure~\ref{fig:system-setup}, weighted by the inverse square of the depicted shell number, resulting in weights from $1$ to $1/16$. This is because the electric field created by an ion decays as $1/r^2$ with the distance $r$. Assuming linear response, the polarizability should be proportional to this local electric field. The weighting also guarantees that charges entering or leaving the cube do not lead to sharp jumps in the dielectric constant, which cause unphysical behavior such as jumps in the total electric field energy. The volume taken into account for the averaging is $(\SI{2.8}{nm})^3$, which for a Bjerrum length of $l_\mathrm{B}\approx\SI{0.7}{nm}$ is roughly the extent over which the electrostatic interactions are significant compared to thermal fluctuations. 

The temporal changes in permittivity lead to additional interactions which have been called spurious~\cite{duncan06a}, but are indeed physical as pointed out by Rottler and Maggs~\cite{rottler11a} and Pasichnyk \etal~\cite{pasichnyk08a}. In our simulations, however, these effects are very small since the algorithm described above only allows smooth and slow changes in the local permittivity.

Initially we examine the counterion distribution around a single rod-like chain of $N=80$ monomers separated by $\SI{0.3}{nm}$ in a cubic box with a side length of $\SI{24}{nm}$ with periodic boundary conditions.
In our conductivity simulations, we vary the monomer concentrations $C$ by placing $M\in \{1,\ldots,25\}$ chains of length $N=30$, $45$, and $60$ in a box of size $(\SI{32}{nm})^3$, depending on the desired polyelectrolyte concentration.
On a modern workstation (Intel i7-5820K CPU, Nvidia GTX 780 Ti GPU), the simulation time for each curve in Figure~4 was around 46 hours. This includes 16 points with 3 different polymer lengths for each, adding up to a total of 48 simulation runs during this time.



\begin{figure}[htbp]
	\centering
	\includegraphics[width=\linewidth]{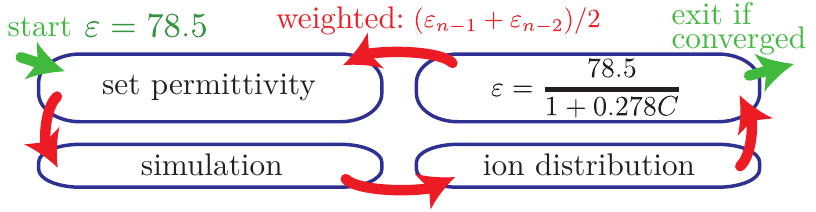}
	\caption{Iterative scheme: Starting with a constant background permittivity, the resulting ion concentrations of successive simulations are mapped to a spatially varying but fixed permittivity distribution, until the scheme converges towards a stable equilibrium.}
	\label{fig:iterative-scheme}
\end{figure}

We first verify the validity of our new approach, in which the dielectric constant dynamically adapts to the local ion concentration. To this end, we use the iterative scheme sketched in Figure~\ref{fig:iterative-scheme} to calculate the counterion distribution around an infinite rod-like polyelectrolyte fixed in space.

\begin{figure}[!htbp]
	\centering
	\includegraphics[width=\linewidth]{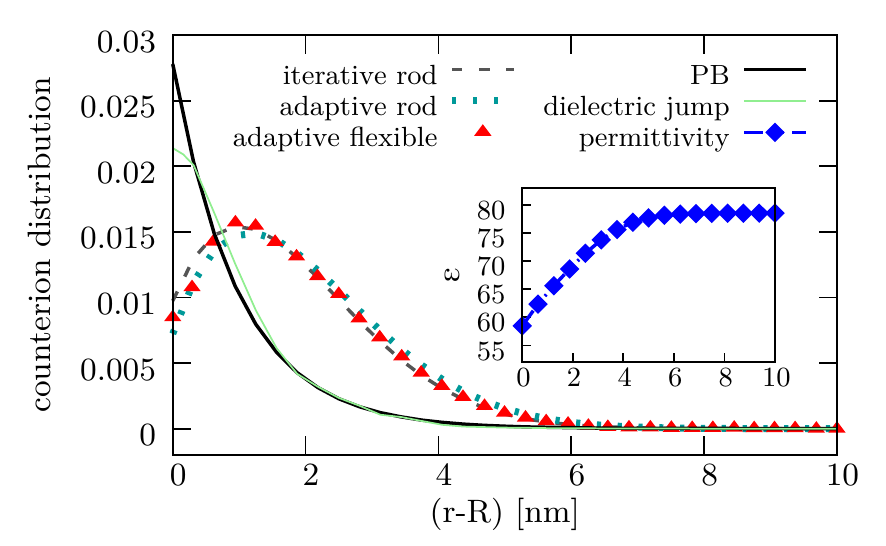}
	\caption{(color online) Counterion distribution around the polyelectrolyte backbone with varying permittivity as a function of the distance from the polyelectrolyte surface. The iterative scheme for a stiff rod (dashed line), the adaptive scheme for a stiff rod (dotted line), and the adaptive scheme for a flexible polymer (red triangles) are almost identical. They differ qualitatively from the analytical Poisson-Boltzmann (PB) solution (thick black line) for a uniform permittivity $\varepsilon=78.5$. The corresponding permittivity $\varepsilon$ (inset) in the iterative case goes from $78.5$ in the bulk to $60$ close to the polyelectrolyte backbone, similar to what is observed near a charged surface~\cite{gongadze12a,gongadze13a}.
	The counterion distribution for a sharp rise in the permittivity from $2$ within the rod to $78.5$ in the fluid (thin light green line) shows only minor deviation from the PB solution.}
	\label{fig:counterion-distribution}
\end{figure}

We start out with a constant permittivity of $\varepsilon = 78.5$ and obtain a counterion distribution from an equilibrium MD simulation. We then map the cylindrically symmetric salt concentration to a local permittivity using equation~\eqref{eq:salt-map} and used it as a fixed permittivity for the subsequent simulation run. A successive under-relaxation scheme equally weighting the two preceding results converges to the counterion distribution in Figure~\ref{fig:counterion-distribution}.
We then apply the new time-dependent adaptive scheme, seen in Figure~\ref{fig:system-setup}, to calculate the local charge concentration and permittivity to a fixed rod-like polyelectrolyte, as well as a fully flexible polyelectrolyte. Figure~\ref{fig:counterion-distribution} displays all three counterion distributions.

The non-monotonicity in the counterion distribution is not predicted by Poisson-Boltzmann (PB) theory with a fixed constant dielectric constant. Note that in the case of a uniform background permittivity there is also a sharp increase in the counterion
density near the polyelectrolyte backbone, however, the increase only extends over a relatively short distance (approximately $0.1\sigma=\SI{0.03}{nm}$, data not shown)
compared to the simulations with a varying dielectric constant (approximately $4\sigma=\SI{1.2}{nm}$). 

The physical basis of the extended initial increase in the counterion concentration when the permittivity is adapted locally is the permittivity gradient in the vicinity of the polyelectrolyte, where there is an increase from approximately $\varepsilon = 60$ at the polymer surface to $\varepsilon = 78.5$ in the bulk (see inset of Figure~\ref{fig:counterion-distribution}). The observed distribution is the combined result of the repulsion of counterions by the permittivity gradient, the electrostatic attraction of the ions to the polyelectrolyte backbone, and thermal fluctuations. Interestingly, there is almost no visible difference in the counterion distribution around the flexible polyelectrolyte and the infinite stiff charged rod, demonstrating that the cell model is a good approximation~\cite{deshkovski01a,antypov06a}.

Our results resemble earlier observations for the counterion distribution around a colloid with spatially varying dielectric background~\cite{fahrenberger13c,ma14a}.
All simulations with varying permittivity also agree qualitatively with atomistic simulations of a similar system using a Kirkwood-Buff based force field for NaCl~\cite{gee11a} in combination with the extended simple point
charge model (SPC/E) water model~\cite{berendsen87a}. The atomistic simulations displayed a similar depletion of counterions very close to the polyelectrolyte backbone and a subsequent rise of the distribution. The excellent agreement between our three simulation setups, and both existing ion distributions for colloids and our atomistic simulations, demonstrates that our method for dynamically adapting the local permittivity produces physically very reasonable results.

In Figure~\ref{fig:counterion-distribution} we have also plotted simulation results where there is a sharp dielectric interface (thin green line) at the surface of the polyelectrolyte,
with a discrete rise from $2$ within the polyelectrolyte to $78.5$ in the fluid. As observed in other studies~\cite{wang15a,nakamura13a}, the discrete change in the dielectric interface
also produces a thickening of the Debye layer. However, the difference in the distribution compared to the
Poisson-Boltzmann (thick black line) result is significantly less than
in our simulations with a permittivity adapted to the local salt
concentration (dotted line). This shows that one not only needs take into account the reduced permittivity
within a polymer, colloid, or charged surface, but also the reduction in the permittivity in the surrounding Debye layer, \ie a gradient in the permittivity.


The structural differences within the EDL in Figure~\ref{fig:counterion-distribution} have little influence on many properties such as the radius of gyration or the polymer diffusion coefficient~\cite{fahrenberger15a-pre}. However, we found that adapting the local dielectric constant significantly impacts the response of the system to an external electric field. This is because the electrokinetic behavior of the system strongly depends on the hydrodynamic and electrostatic friction between the polyelectrolyte backbone and its counterions, and is thus very susceptible to changes within the EDL.

\begin{figure}[!htbp]
	\centering
	\includegraphics[width=\linewidth]{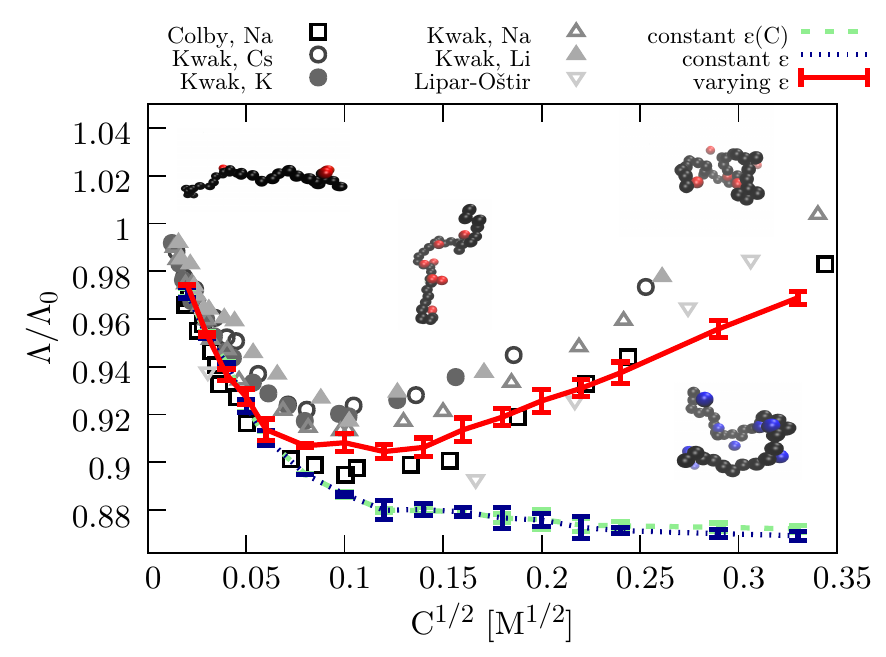}
	\caption{(color online) The rescaled equivalent conductivity $\Lambda/\Lambda_0$ over monomer concentration $\sqrt{C}$ for simulations with constant permittivity (green dashed, blue dotted) and locally varying (red solid) permittivity. Experimental data (gray symbols) from Kwak and Hayes~\cite{kwak75a}, Colby~\etal{}~\cite{colby97a}, and Lipar-O\v{s}tir~\etal{}~\cite{liparostir09a} is reproduced with locally varying permittivity, while we observe a qualitative difference for a constant dielectric background. The reason for this is a significant drop in the dielectric constant around the polyelectrolyte backbone, as seen in Figure~\ref{fig:epsilon-cci}, and the subsequent repulsion of counterions from the polymer.}
	\label{fig:conductivities-concentration}
\end{figure} 

We simulated the equivalent conductivity $\Lambda$ (the conductivity over the polyelectrolyte concentration) of a salt-free solution of polyelectrolytes as a function of the monomer concentration $C$ as plotted in Figure~\ref{fig:conductivities-concentration} together with experimental data. The equivalent conductivity is simply:
\begin{equation}
	\Lambda = \frac{\vec{j}}{EC} ,
	\label{eq:equiv-cond}
\end{equation}
where $\vec{j}$ is the current density, $E$ is the applied electric field, and $C$ is the monomer concentration.
The conductivities have been additionally normalized by an extrapolated $\Lambda_0 (C=0)$, since the hydrodynamic radius of the specific ions strongly influences the diffusion and conductivity even at infinite dilution. 
We performed two sets of simulations with a constant background permittivity, one with $\varepsilon_r=78.5$ and one where the ion concentration in the simulation box was substituted into equation~\ref{eq:salt-map} to determine the background permittivity. 
Both sets of simulations with a constant dielectric background show a monotonic decrease in the equivalent conductivity, in good agreement with scaling theories which ignore the effect of dielectric contrast~\cite{manning69a,schmitt73a,gonzalezmozuelos95a,colby97a,dobryin95a,cametti14a}. 
More importantly, our simulations that adapt the permittivity to the local ion concentration quantitatively reproduce the unexpected rise in conductivity at high salt concentrations.

\begin{figure}[!htbp]
	\centering
	\includegraphics[width=\linewidth]{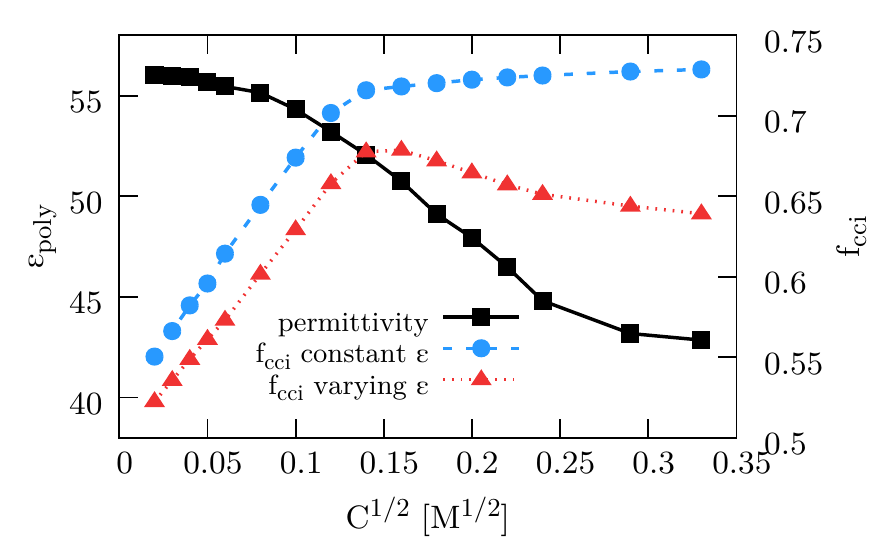}
	\caption{(color online) The local permittivity $\varepsilon_\mathrm{poly}$ around the polymer backbone (black squares), and the fraction of condensed counterions $f_\mathrm{cci}$ for constant dielectric background (blue circles) and varying local permittivity (red triangles) as a function of the monomer concentration. The maximum in the fraction of condensed counterions is due to the sharp decrease in the permittivity with increasing monomer concentration and occurs at the same concentration as the minimum in the equivalent conductivity in Figure~\ref{fig:conductivities-concentration}. }
	\label{fig:epsilon-cci}
\end{figure}

From the snapshots in Figure~\ref{fig:conductivities-concentration} at monomer concentrations of $\SI{1}{mM}$, $\SI{10}{mM}$, and $\SI{100}{mM}$, we see a coiling of the polyelectrolytes (black spheres) as well as a decrease in the fraction of condensed counterions (red spheres) at high concentrations for the simulations including dielectric variations (top right snapshot). The simulations assuming a constant dielectric background (bottom right snapshot) resulted in a larger number of condensed counterions (blue spheres). 
To quantify the fraction of condensed counterions, we used the criterion suggested by L.~Belloni~\cite{belloni84a} and M.~Deserno~\cite{deserno00a}. In Figure~\ref{fig:epsilon-cci}, we plot the fraction of condensed counterions $f_\mathrm{cci}$ as a function of the monomer concentration $C$.

While $f_\mathrm{cci}$ monotonically increases for simulations with a constant dielectric background, we observe a maximum and subsequent decrease when we adapt the dielectric constant to the local ion concentration. The maximum occurs at the same monomer concentration as the minimum in conductivity in Figure~\ref{fig:conductivities-concentration}. The decrease in $f_\mathrm{cci}$ results in a higher effective charge of the polyelectrolytes, which both increases their mobility and the net charge they carry with them. In addition, there is an increase in the number of free counterions contributing to the overall conductivity. This is why the maximum in the fraction of condensed counterions results in a minimum in the equivalent conductivity of the solution.

We average the dielectric permittivity within $\SI{1.4}{nm}$ of the polyelectrolyte backbone and plot the results as black squares in Figure~\ref{fig:epsilon-cci}. We find that the reason for the drop in $f_\mathrm{cci}$ is a decrease in dielectric permittivity and therefore a steeper gradient in $\varepsilon$, resulting in a stronger dielectric repulsion from the backbone. The decrease in $\varepsilon$ is enhanced due to the coiling of the chains at high polyelectrolyte concentrations as seen in the snapshots in Figure~\ref{fig:conductivities-concentration}. This increases the density of ions within the polymer coil, which leads to the large decrease in the local permittivity around the polyelectrolyte backbone. 

The dependence of the conductivity on the monomer concentration is largely independent of the specific system being studied. This can be seen by the comparison of three different experimental data sets in Figure~\ref{fig:conductivities-concentration}. Colby~\etal{}~\cite{colby97a} measured the equivalent conductivity poly(styrenesulfonate) (PSS) with Na$^+$ counterions at 296 K with a molecular weight of $M=\num{1.2d6}$. In contrast, Kwak and Hayes~\cite{kwak75a} used poly(styrenesulfonic acid) (PSA) with Li$^+$, Na$^+$, K$^+$, Cs$^+$ at 298 K with a molecular weight of $M=\num{5.6d5}$. Finally, Lipar-O\v{s}tir~\etal{}~\cite{liparostir09a} measured the equivalent conductivity of poly(anetholesulfonic acid) (PAS) with H$^+$ measuring seven sets at temperatures between $278$ and $\SI{308}{K}$. Despite the different counterion types, temperatures, polyelectrolytes, and polyelectrolyte lengths, all experimental data sets collapse onto our simulations that include variations of the local dielectric constant. This demonstrates that our model captures the essential physics.

Our results demonstrate that the inclusion of local variations in the dielectric permittivity in coarse-grained MD simulations strongly influences the static and dynamic properties of charged macromolecules in aqueous solutions and should be taken into account in the modeling process. We expect that dielectric contrast is a key factor in determining the electrokinetic properties of soft matter systems, such as colloid and polyelectrolyte electrophoresis; electroosmotic flow; and induced charge electroosmosis and electrophoresis, and that our algorithm can be applied successfully to all these systems.

\section*{Acknowledgements}

We thank the DFG for support through the SFB 716 and HO 1108/22-1, the German Ministry of Science and Education (BMBF) for support under grant 01IH08001, and the Volkswagen foundation. We would also like to thank Joost de Graaf, Shervin Raafatnia and Gary Davies for helpful discussion.

\bibliographystyle{apsrev4-1}
\bibliography{fahrenberger15b}

\end{document}